\documentclass{aa}  
\usepackage{graphicx}
\usepackage{txfonts}
\begin{document}
   \title{}

   \subtitle{INTEGRAL and \emph{Swift}  observations  of the supergiant fast X-ray transient  AX~J1845.0$-$0433=IGR~J18450$-$0435}

   \author {V. Sguera\inst{1}, A. J. Bird\inst{1}, A. J. Dean\inst{1}, A. Bazzano\inst{2}, P. Ubertini\inst{2},  R. Landi\inst{3}, 
  A. Malizia\inst{3}, E. J. Barlow\inst{1}, D. J. Clark\inst{1},
   A. B. Hill\inst{1}, M. Molina\inst{1}}

   \offprints{sguera@astro.soton.ac.uk 
}
   \institute{ School of Physics and Astronomy, University of Southampton, Highfield, SO17 1BJ, UK \and 
   IASF/CNR, via Fosso del Cavaliere 100, 00133 Roma, Italy \and 
   IASF/CNR, via Piero Gobetti 101, I-40129 Bologna, Italy
             }

   \date{Received 8 August 2006 / accepted 25 October 2006}


  \abstract
   {AX~J1845.0$-$0433 was discovered by ASCA in 1993 during fast outburst activity
    characterized by several flares on short timescales. Up to now, the source was not detected again by any X-ray  mission.
    Its optical counterpart is suggested to be an O9.5I supergiant star, which is the only remarkable object found inside the ASCA error box.}
   {To detect and characterize new fast outbursts of AXJ1845.0$-$0433 and confirm its supergiant HMXB nature, using INTEGRAL and archival \emph{Swift} XRT observations.}
   {We performed an analysis of INTEGRAL IBIS and JEM--X data with OSA 5.1 as well as an analysis of archive \emph{Swift} XRT data}
   {We report on fast flaring activity from the source on timescales of a few tens of minutes, the first to be reported since its discovery in 1993. 
    Two outbursts have been detected by INTEGRAL (Apr 2005 and Apr 2006) with similar peak fluxes and peak luminosities 
     of $\sim$ 80 mCrab and 9.3$\times$10$^{35}$ erg s$^{-1}$ (20--40 keV), 
    respectively.  Two other outbursts were detected by \emph{Swift} XRT on Nov 2005 and Mar 2006. 
    The refined  \emph{Swift} XRT position of AX~J1845.0$-$0433 confirms its association with  the supergiant star previously proposed as optical counterpart.}
   {Our INTEGRAL and \emph{Swift} results fully confirm the supergiant HMXB nature of AX~J1845.0$-$0433 which 
    can therefore be classified as a supergiant fast X-ray transient. Moreover they provide for 
    the first time evidence of its recurrent fast transient behaviour.}

   \maketitle

\section{Introduction}

The X-ray transient AX~J1845.0$-$0433 was discovered in the Scutum arm region on 18 October 1993 during an ASCA  
observation lasting $\sim$ 16 hours (Yamauchi et al. 1995).
During the initial $\sim$ 9 hours,  the source was in a very faint quiescent state with a 0.7--10 keV  flux of $\sim$ 3$\times$10$^{-12}$ erg cm$^{-2}$ s$^{-1}$, 
then suddenly it flared up reaching a peak flux of   $\sim$ 10$^{-9}$ erg cm$^{-2}$ s$^{-1}$  in less than 20 minutes. 
Subsequently the source was characterized by several other peaks lasting a few tens of minutes, until the end of the observation. 
Optical and infrared  measurements of the ASCA error circle (1 arcmin radius) of AX~J1845.0$-$0433 were performed by Coe et al. (1996).
The only object of interest is an O9.5I supergiant star which was proposed by Coe et al. (1996) as the optical counterpart.
Its estimated distance was $\sim$ 3.6 kpc, consistent with that derived from ASCA X-ray measurements. 
However Coe et al. (1996) pointed out that the error on the distance could be large,
mainly  because of the uncertainty in the reddening law. With these assumptions, 
the quiescent and the peak luminosity (0.7--10 keV) of the source measured by ASCA were  
4.6$\times$10$^{33}$ erg s$^{-1}$ and 1.5$\times$10$^{36}$ erg s$^{-1}$, respectively.
The ASCA X-ray spectrum of the flare state was fitted by an absorbed  power law 
with $\Gamma$$\sim$1 and N$_{H}$=3.6$\pm$0.3$\times$10$^{22}$cm$^{-2}$; no coherent pulsation was found in the range from 125 ms to 4096 s (Yamauchi et al. 1995).

Since the ASCA discovery, no further outburst  activity  has been reported from the source by any X-ray mission.  
The fast X-ray transient behaviour of AX~J1845.0$-$0433 as well as its likely, but not entirely definitive,
association with a supergiant star, suggest its classification as a possible member of the recently newly discovered class 
of Supergiant Fast X-ray Transients: SFXTs (Negueruela et al. 2005, Sguera et al. 2005, Sguera et al. 2006).
They are supergiant high mass X-ray binaries (SGXBs), which most of the time are undetectable having typical quiescence 
luminosities  $\leq$ 10$^{32}$--10$^{33}$  erg s$^{-1}$. Occasionally they undergo  
fast X-ray transient activity  lasting typically less than a day  and characterized 
by several fast flares with timescales of a few tens of minutes and typical peak luminosities of  $\sim$ 10$^{36}$ erg s$^{-1}$. 

Recently, Halpern et al. (2006) noted that AX~J1845.0$-$0433 is very likely the same object as IGR~J18450$-$0435.
Using the precise position of its optical counterpart, AX~J1845.0$-$0433 is located 2$^{'}$.3 off the IBIS position 
of IGR~J18450$-$0435 reported by Bird et al. (2006), slightly outside the ISGRI error circle.
IGR~J18450$-$0435 was firstly discovered by INTEGRAL  during a survey  of the Sagittarius arm tangent region  in the spring 2003
(Molkov et al. 2004) at an average  flux of 1.5$\pm$0.3 mCrab (18--60 keV). 

Here we report on the renewed fast outburst activity detected by INTEGRAL and \emph{Swift} from AX~J1845.0$-$0433, the first 
to be revealed since its discovery in 1993.

\begin{figure}[t!]
\centering
\includegraphics[width=8cm,height=6.5cm]{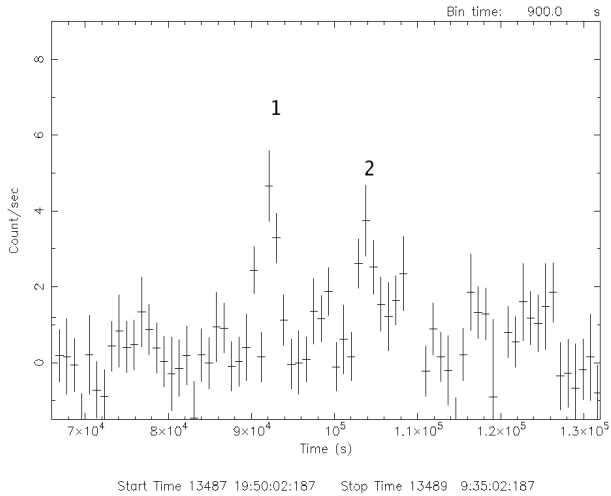}
\caption{ISGRI light curve (20--40 keV) of a newly discovered outburst of IGR~J18450$-$0433 on 28 April 2005}
\end{figure}

\section{INTEGRAL observations and results}

The INTEGRAL gamma-ray observatory carries three co-aligned coded mask telescopes: the imager IBIS (Ubertini et al. 2003),
the spectrometer SPI (Vedrenne et al. 2003) and the X-ray monitor JEM--X (Lund et al. 2003). In this work only results from
IBIS/ISGRI (Lebrun et al. 2003) and JEM--X will be presented.  
The reduction and analysis of the IBIS/ISGRI and the JEM--X data have been performed 
using the INTEGRAL Offline Scientific Analysis (OSA) v.5.1.
INTEGRAL observations are typically divided into short pointings (Science Window, ScW) of $\sim$ 2000 s duration.

We performed an analysis at the ScW level of the deconvolved ISGRI shadowgrams searching for outburst activity from IGR~J18450$-$0435.
In most of the ScWs, the source was  well below a 5$\sigma$ detection level (20--40 keV), and only in two  
ScWs in revolution 310 and one ScW in revolution 429 was it significantly detected at $\sim$ 6$\sigma$ level (20--40 keV).
Our analysis showed that these detections corresponded to fast flares reaching their peak in a few tens of minutes and then dropping on the same timescale.
In section 2.1 and 2.2 we report on INTEGRAL results of each outburst.

\subsection{INTEGRAL results on the first outburst activity}

\begin{figure}
\centering
\includegraphics[width=6cm,height=4cm]{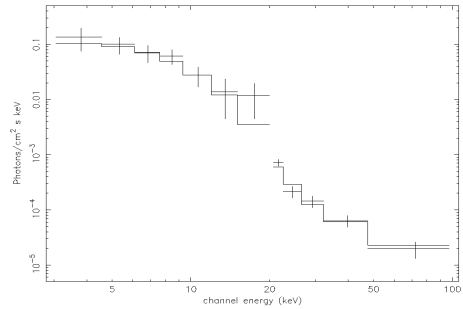}
\caption{Unfolded BMC broad band spectrum (3--100 keV) of AX~J1845.0$-$0433 during the flare that occurred on 28 April 2005.}
\end{figure}

Figure 1 displays the 20--40 keV light curve of IGR~J18450$-$0435. As we can see, the source underwent two fast flares 
(No. 1 and No. 2 in figure 1) with timescales of a few tens 
of minutes on 28 April 2005 at $\sim$ 19:40 UTC.
The first flare reached a peak flux of $\sim$ 60 mCrab (20--40 keV). Assuming a distance of 3.6 kpc, 
the 20--40 keV peak luminosity was $\sim$ 7$\times$10$^{35}$ erg s$^{-1}$, in agreement with typical outburst luminosities 
of SFXTs (Sguera et al. 2006). After the second flare, the source seemed to show no further X-ray flaring activity.
However the light curve is unfortunately truncated at the end because the source went outside  the IBIS FOV.
It is not possible to constrain the overall duration of the outburst activity  and we cannot exclude the possibility that further flares took place.
The INTEGRAL observation coverage since the source turned on up to the end of the light curve is $\sim$~11 hours.
A search for subsequent flaring activity, when the source was again in the IBIS FOV $\sim$ 1.4 days later, did not give any positive results.

The first flare (No. 1 in figure 1) was also covered with JEM--X data.
The 3--10 keV peak flux and peak luminosity were $\sim$ 32 mCrab and $\sim$ 7.4$\times$10$^{35}$ erg s$^{-1}$, respectively. 
The combined JEM--X and ISGRI spectrum (3--100 keV) during the first flare
was reasonably  fitted by a single power law ($\chi^{2}_{\nu}$=1.12, 156 d.o.f.) with $\Gamma$=2.5$^{+0.6}_{-0.5}$.
The fit improved adding a black body to the power law ($\chi^{2}_{\nu}$=1.02, 154 d.o.f.), being the best fit parameters kT=2$^{+0.7}_{-0.5}$ and
$\Gamma$=1.7$^{+0.7}_{-0.8}$. To account for a cross-calibration mismatch between the two instruments we have introduced a constant in the fit, 
which when left free to vary provides a value of 0.02$^{+0.08}_{-0.016}$.
An equally good  fit ($\chi^{2}_{\nu}$=1.02, 154 d.o.f.)  was also achieved using a Bulk Motion Comptonization  model BMC (see figure 2).

\begin{figure}[t!]
\centering
\includegraphics[width=8cm,height=6.5cm]{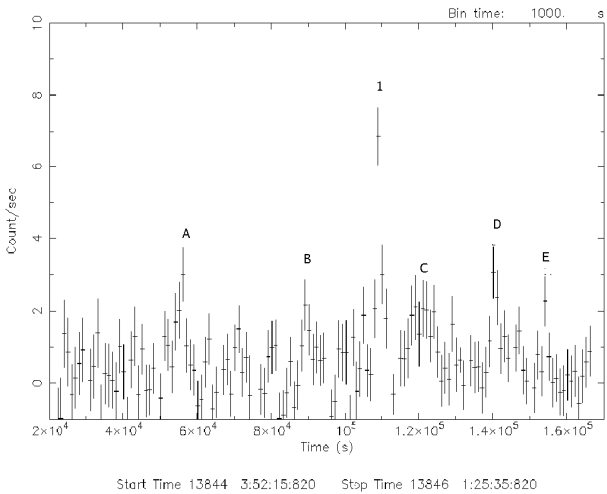}
\caption{ISGRI light curve (20--40 keV) of a newly discovered outburst of AX~J1845.0$-$0433 which occurred on 20 April 2006.}
\end{figure}

Our analysis of this first flare enabled the determination of an ISGRI position for IGR~J18450$-$0435
(RA=18$^{h}$ 45$^{m}$ 03$^{s}$.3 DEC=-04$^{\circ}$ 34$^{'}$ 05$^{''}$.5, (J2000), error radius of 2.$^{'}$4)
which is located 0$^{'}$.5 from the optical counterpart of AX~J1845.0$-$0433. 
This fully confirms that AX~J1845.0$-$0433 and  IGR~J18450$-$0435 are indeed the same source.

\subsection{INTEGRAL results on the second outburst activity} 

Figure 3 displays  the 20--40 keV light curve of IGR~J18450$-$0435 using recent Core Program data in revolution 429. 
As we can note, it strongly resembles the light curve of the previous flaring activity reported in section 2.1. Most of the time the source did
not show any relevant outburst activity, with a count rate less than 1 count/sec, however several very quick flares (labeled from A to E in figure 3)
appeared on a timescale of tens of minutes. Their peak flux never reached a value greater than $\sim$ 30 mCrab (20--40 keV). 
Moreover a noticeable very fast and strong flare (No. 1 in figure 3)
occurred on $\sim$ 9:30 UTC 20 April 2006. It reached a peak flux of $\sim$ 80 mCrab or 6$\times$10$^{-10}$ erg cm$^{-2}$ s$^{-1}$
(20--40 keV) in $\sim$ 30 minutes and then it dropped with the same timescale. The 20--40 keV peak luminosity was $\sim$ 9.3$\times$10$^{35}$ erg s$^{-1}$. 
These values are slightly greater than the previous ones relating to the flaring activity detected by INTEGRAL about one year before (section 2.1).
The light curve in figure 3, pertaining to rev 429, is truncated at the beginning and at the end. A search for more flaring activity before and after
this revolution did not give any results.
We extracted an  ISGRI spectrum during this fast flare (20--100 keV), the best fit ($\chi^{2}_{\nu}$=1.18, 24 d.o.f.)
was provided applying a single power law with $\Gamma$=2.9$^{+0.9}_{-0.7}$. However a BMC model  also provided a good description to the data
($\chi^{2}_{\nu}$=1.2, 22 d.o.f.)

\section{\emph{Swift} observations and results}

\begin{figure}[t!]
\centering
\includegraphics[width=7cm,height=5cm]{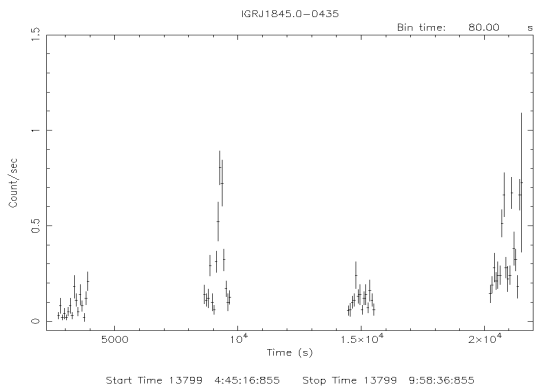}
\caption{\emph{Swift} XRT light curve of  AX~J1845.0$-$0433 (0.2--10 keV) during the observation on 5 March 2006}
\end{figure}

\begin{figure}[h!]
\centering
\includegraphics[width=7cm,height=5cm]{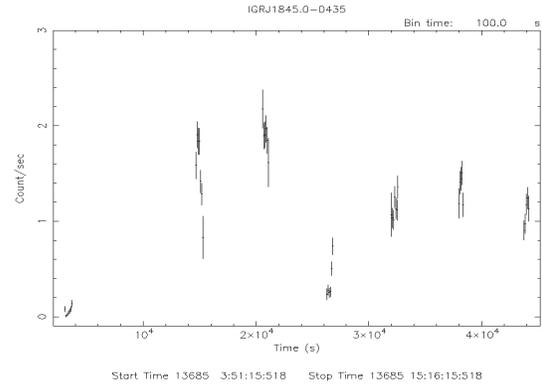}
\caption{\emph{Swift} XRT light curve of  AX~J1845.0$-$0433 (0.2--10 keV) during the observation on 11 November 2005}
\end{figure}

In this section we present X-ray observations acquired with the XRT 
(X-ray Telescope, 0.2--10 keV) on board the 
\emph{Swift} satellite (Gehrels et al. 2004). A search of the XRT data archive revealed that \emph{Swift} carried out  2 
observations of IGR~J18450$-$0433 on 11 November 2005 and 5 March 2006.
Unfortunately in both observations the source was outside the IBIS  FOV so we 
had no simultaneous ISGRI data. The XRT collected data for a total exposure 
time of 4.6 ks and 4.8 ks, respectively. 
The data reduction was performed using XRTDAS v. 2.4 standard data 
pipeline package ({\sc xrtpipeline} v. 0.10.3), in order to produce 
screened event files. All data were extracted only in the Photon Counting 
(PC) mode, 
adopting the standard grade filtering (0--12 
for PC) according to the XRT nomenclature. Events for spectral analysis 
were extracted within a circular region of radius 20$^{\prime \prime}$, 
which encloses about 90\% of the PSF at 1.5 keV (Moretti et al. 2004) 
centered on the source position.
The background was extracted from various source-free regions close to the 
X-ray source of interest using both circular/annular regions with 
different radii, in order to ensure an evenly sampled background.

\begin{table*}[t!]
\begin{center}
\caption {Summary of INTEGRAL, \emph{Swift} and ASCA  observations of outbursts of AX~J1845.0$-$0433}
\begin{tabular}{cccccccc}
\hline
\hline
No. & Observation & Date  &  energy band & peak flux       & peak luminosity  & photon index  & N$_{H}$ \\
    &             &       &  (keV)            &(erg cm$^{-2}$ s$^{-1}$) & (erg s$^{-1}$)   & (power law) &  (cm$^{-2}$)\\
\hline
1  & ASCA           & 18 Oct 1993 & 0.7--10 & 1$\times$10$^{-9}$ &  1.5$\times$10$^{36}$ & 1$^{+0.07}_{-0.07}$ & 3.6$\pm$0.3$\times$10$^{22}$ \\
2  & INTEGRAL/ISGRI & 28 Apr 2005 & 20--40  & 4.5$\times$10$^{-10}$  & 7$\times$10$^{35}$ &  2.5$^{+0.6}_{-0.5}$ \\
3  & INTEGRAL/JEM--X & 28 Apr 2005 & 3--10  &  4.8$\times$10$^{-10}$  & 7.4$\times$10$^{35}$  & \\
4  & \emph{Swift} XRT & 11 Nov 2005 & 0.2--10  & 2.3$\times$10$^{-10}$  & 3.6$\times$10$^{35}$  &   0.75$^{+0.1}_{-0.1}$ & 1.6$\pm$0.18$\times$10$^{22}$\\  
5  & \emph{Swift} XRT & 5 Mar 2006 & 0.2--10  & 1.1$\times$10$^{-10}$  &  2$\times$10$^{35}$ &   0.85$^{+0.3}_{-0.3}$  & 2.3$\pm$0.7$\times$10$^{22}$ \\
6  &  INTEGRAL/ISGRI & 20 Apr 2006 &  20--40  & 6$\times$10$^{-10}$ & 9.3$\times$10$^{35}$ &  2.9$^{+0.9}_{-0.7}$     \\
\hline
\end{tabular}
\end{center}
\end{table*} 

Figures 4 and 5 show the 0.2--10 keV  XRT light curve of IGR~J18450$-$0433 during the observations performed on 5 March 2006 and 11 November 2005, respectively.
We note that the temporal coverage was not continuous,  there were significant gaps in both the light curves due to visibility constraints.
Nevertheless the fast flaring behaviour of the source was evident, especially during the observation performed on 5 March 2006 (figure 4). 
On this occasion, initially the source displayed a very low count rate. 
Subsequently, it underwent a very fast flare reaching in only 4 minutes a peak flux and a peak luminosity
of $\sim$  1.1$\times$10$^{-10}$ erg cm$^{-2}$ s$^{-1}$ and  $\sim$ 2$\times$10$^{35}$ erg s$^{-1}$, respectively (0.2--10 keV).
Then the flux dropped again to a very low level with the same fast timescale.
The spectrum (0.2--8 keV) extracted during the observation 5 March 2006 was  well fitted by an absorbed power law 
($\chi^{2}_{\nu}$=0.63, 41 d.o.f) with  $\Gamma$=0.85$\pm$0.3 and N$_{H}$=2.3$\pm$0.7$\times$10$^{22}$ cm$^{-2}$.
However a good fit was also  provided by a black body model ($\chi^{2}_{\nu}$=0.7, 42 d.o.f) with kT=2.4$\pm$0.2. 
The average 0.2--8 keV flux was $\sim$2$\times$10$^{-11}$ erg cm$^{-2}$ s$^{-1}$.

In the light curve in figure 5 (11 November 2005), initially the source was characterized by a count rate less than 0.1 count/s,  then it flared up quickly 
reaching a peak flux and peak luminosity of $\sim$ 2.3$\times$10$^{-10}$ erg cm$^{-2}$ s$^{-1}$ and  
$\sim$ 3.6$\times$10$^{35}$ erg s$^{-1}$ respectively (0.2--10 keV).
The spectrum (0.2--10 keV) was best fitted by an absorbed power law ($\chi^{2}_{\nu}$=1.28, 166 d.o.f.) 
with $\Gamma$=0.75$\pm$0.1 and N$_{H}$=1.6$\pm$0.18$\times$10$^{22}$ cm$^{-2}$.
The latter is in agreement with the galactic absorption along the line of sight (N$_{H}$=1.58$\times$10$^{22}$ cm$^{-2}$). 
A simple power law or thermal models, such as blackbody or  bremsstrahlung, instead provided very poor fits.  
The 0.2--10 keV average flux was $\sim$ 1.3$\times$10$^{-10}$ erg cm$^{-2}$ s$^{-1}$, which is an order of magnitude higher than that during the other observation.

Our \emph{Swift} analysis provided a very accurate position of AX~J1845.0$-$0433 (RA=18$^{h}$ 45$^{m}$ 01.$^{s}$9 DEC=-04$^{\circ}$ 33$^{'}$ 57$^{''}$.6)
which is located 4$^{''}$.7 from the supergiant star proposed by Coe et al. (1996) as its optical counterpart.
This fully confirms the supergiant HMXB nature of  AX~J1845.0$-$0433.

\section{Conclusions}

To date, the only X-ray detection of  AX~J1845.0$-$0433 in outburst dates back to 1993. 
No X-ray mission provided a regular
monitoring of the Scutum arm region, which is essential to detect fast X-ray transient activity from a source like AX~J1845.0$-$0433.
The INTEGRAL satellite is regularly monitoring the Scutum arm region performing cycles of observations
with an exposure time that is $\sim$ 10 days. We reported  on renewed fast outburst activity from AX~J1845.0$-$0433
discovered  by INTEGRAL and also on results of two \emph{Swift} observations from archival data.
Table 1 provides a summary of the characteristics of all outbursts from  AX~J1845.0$-$0433 detected to date.

As for the newly discovered outbursts by INTEGRAL, 
they occurred on 28 April 2005 and 20 April 2006. The strongest one was characterized by a 20--40 keV peak 
flux and peak luminosity of $\sim$ 80 mCrab (6$\times$10$^{-10}$ erg cm$^{-2}$ s$^{-1}$)
and $\sim$ 9$\times$10$^{35}$ erg s$^{-1}$, respectively. In both of them, the source displayed 
a kind of pre and post flaring activity characterized by several other small flares having the same fast timescale but a smaller peak 
flux which never reached a value greater than 30 mCrab (20--40 keV). The combined JEM--X/ISGRI  spectrum (3--100 keV) was 
well described by a power law plus black body model  or by a Comptonized model (BMC). 

From \emph{Swift} XRT data analysis, we fully confirmed that the supergiant star O9.5I is  
the optical counterpart of AX~J1845.0$-$0433 as result of the very accurate \emph{Swift} position of AX~J1845.0$-$0433, located 4$^{''}$.7 from the star. 
During the first \emph{Swift} observation (11 November 2005), the average flux of the source ($\sim$ 1.3$\times$10$^{-10}$ erg cm$^{-2}$ s$^{-1}$) 
was an order of magnitude higher than that during  the second observation 
(5 March 2006). On the contrary, the spectrum was very similar in both observations being well fitted by an absorbed power law with $\Gamma$ $\sim$ 0.75--0.85 and 
N$_{H}$ $\sim$ 2$\times$10$^{22}$ cm$^{-2}$. The latter is similar to that measured by ASCA in 1993 (see table 1).
We point out that the galactic absorption along the line of sight is 1.58$\times$10$^{22}$ cm$^{-2}$.

The four newly discovered fast outbursts reported in this paper provided evidence of  recurrent fast X-ray transient behaviour of  
the supergiant high mass X-ray binary AX~J1845.0$-$0433. Its fast flaring activity sets it   
apart from  classical  SGXBs and allows  its classification as member of the recently  discovered class 
of Supergiant Fast X-ray Transients: SFXTs (Negueruela et al. 2005, Sguera et al. 2005, Sguera et al. 2006).
To date,  AX~J1845.0$-$0433 is the sixth such source to be detected by INTEGRAL.
Its  flaring behaviour is very similar to that of the others 5 SFXTs, with the same 
fast timescale, peak-flux and peak-luminosity. 
Fast X-ray activity has been detected from the source in ASCA, \emph{Swift} and INTEGRAL observations lasting only a few hours.
Since INTEGRAL is regularly monitoring the Scutum arm region, further IBIS detections  of AX~J1845.0$-$0433 in outburst
are not  unexpected. This could allow us to search for recurrence time and in turn further insights into the system geometry.

\begin{acknowledgements}
We thank the anonymous referee for very useful comments which helped us to improve the paper.
This research has been supported by  University of Southampton School of Physics and Astronomy. 
AB, PU, AM, RL acknowledge the ASI financial support via grant I/R/046/04.

\end{acknowledgements}

\end{document}